%% file: paper_draft.tex
\documentclass[conference, comsoc]{IEEEtran}

\usepackage[T1]{fontenc}
\usepackage{cite}
\usepackage[pdftex]{graphicx}
\graphicspath{{./figures/}}
\usepackage{empheq} % autoloads mathtols and amsmath
\usepackage{cases}
\usepackage[cmintegrals]{newtxmath}
\usepackage{bm}
\usepackage{algorithm}
\usepackage[noend]{algpseudocode}  % algorithmicx package is loaded by this package
\usepackage[caption=false,font=footnotesize]{subfig}
\usepackage{dblfloatfix}
\ifCLASSOPTIONcaptionsoff
  \usepackage[nomarkers]{endfloat}
 \let\MYoriglatexcaption\caption
 \renewcommand{\caption}[2][\relax]{\MYoriglatexcaption[#2]{#2}}
\fi
\usepackage{url}
\usepackage{color}
\usepackage[pdfencoding=auto, psdextra]{hyperref}
\DeclareMathAlphabet{\pazocal}{OMS}{zplm}{m}{n}
\newcommand{\Ob}{\pazocal{O}}

\begin{document}
	
\title{Non-uniform array design for robust LoS MIMO via convex optimization}

\author{\IEEEauthorblockN{Michail Palaiologos\textsuperscript{$\ast$\dag}, Mario H. Cast\~aneda Garc\'ia\textsuperscript{$\ast$}, Anastasios Kakkavas\textsuperscript{$\ddagger$}, Richard A. Stirling-Gallacher\textsuperscript{$\ast$}\\
and Giuseppe Caire\textsuperscript{\dag}}
\IEEEauthorblockA{\textsuperscript{*}Munich Research Center, Huawei Technologies D{\"u}sseldorf GmbH, 80992 Munich, Germany\\
\IEEEauthorblockA{\textsuperscript{$\ddagger$} School of Computation, Information and Technology, Technical University of Munich, 80333 Munich, Germany\\ 
\IEEEauthorblockA{\textsuperscript{\dag}Communications and Information Theory Group, Technical University of Berlin, 10587 Berlin, Germany\\
Emails: \{\href{mailto:michail.palaiologos@huawei.com}{michail.palaiologos}, 
\href{mailto:mario.castaneda@huawei.com}{mario.castaneda},
\href{mailto:richard.sg@huawei.com}{richard.sg}\}@huawei.com, 
\href{mailto:tasos.kakkavas@tum.de}{tasos.kakkavas@tum.de},
\href{mailto:caire@tu-berlin.de}{caire@tu-berlin.de}}}}}

\maketitle
	
\begin{abstract}
The array design problem of multiple-input multiple-output (MIMO) systems in a line-of-sight (LoS) transmit environment is examined. As uniform array configurations at the transmitter (Tx) and receiver (Rx) are optimal at specific transmit distances only, they lead to reduced spectral efficiency over a range of transmit distances. To that end, the joint design of non-uniform Tx and Rx arrays towards maximizing the minimum capacity of a LoS MIMO system across a range of transmit distances is investigated in this paper. By introducing convex relaxation, the joint Tx and Rx array design is cast as a convex optimization problem, which is solved in a iterative manner. In addition, we also implement a local search to obtain a refined solution that achieves an improved performance. It is shown that the non-uniform configurations designed with our proposed approach outperform uniform and non-uniform array designs of the literature in terms of capacity and/or complexity.
\end{abstract}

\IEEEpeerreviewmaketitle

\section{Introduction}\label{SecI}
A shift towards millimeter wave and sub-THz frequency bands is one of the  main enablers for satisfying the higher data rates required in future wireless communication systems \cite{9170653}. However, as at higher frequencies the channel may be dominated by the line-of-sight (LoS) path, the deployment of multiple antennas can result in rank deficient MIMO channels, thus eluding an increase of the spatial degrees of freedom \cite{7370753}.

Still, it has been demonstrated that, if the Tx and Rx arrays are properly designed, spatial multiplexing gain can be extracted, even if the channel is dominated by the LoS path. Specifically, optimum antenna placement has been proposed for uniform linear arrays (ULAs) \cite{1175470}, uniform planar arrays (UPAs) \cite{1543276} and uniform circular arrays (UCAs) \cite{9363635}. A major shortcoming of prior uniform designs is that the proposed arrays are optimized for a specific \textit{transmit distance} (distance between the Tx and Rx array). As such a design does not consider the performance at other transmit distances, substantial capacity fluctuations and, thus, decreased performance occurs when operating over a wide range of distances.

To circumvent this, non-uniform array designs have been proposed \cite{5425310, 8430512, 8955847, 9049013, 7546944, electronics6030057}. By assuming fixed arrays consisting of non-uniformly spaced antennas at the Tx and/or Rx, reduced capacity fluctuations and, hence, improved performance can be achieved over a range of transmit distances. However, due to the complexity of the LoS MIMO array design problem, which is highly non-convex, most prior works rely on brute force exhaustive search (ES) \cite{5425310, 8430512}, stochastic optimization techniques, such as genetic algorithms (GAs) \cite{8955847, 9049013}, and on numerical results. Specifically, by applying ES and assuming identical antenna locations at the Tx and Rx, non-uniform linear arrays (NULAs) are derived for $4 \times 4$ systems to maximize the range of transmit distances over which capacity remains above a threshold \cite{5425310} and the minimum capacity over a given range of transmit distances \cite{8430512}. In \cite{8955847} a non-uniform planar array is obtained via a GA by maximizing the mean minus the standard deviation of the capacity over a range of distances. In \cite{9049013}, a NULA for a LoS massive MIMO system is obtained through a GA for maximizing the user sum rate.

In \cite{7546944} the channel matrix eigenvalues are expressed as a function of the Tx and Rx antenna locations. In this way, optimum Tx and Rx NULA configurations are obtained, so that a minimum number of spatial streams is supported. In \cite{electronics6030057}, a multi-user LoS massive MIMO scenario is considered, where the antenna locations of the Tx NULA are obtained by Chebyshev polynomials. It is shown that the average condition number is greatly improved compared to a system with ULAs. \cite{7546944} and \cite{electronics6030057} are tailored to the design of NULAs and cannot be applied in the design of arrays of arbitrary geometries. In all works, either identical Tx and Rx non-uniform configurations are considered for single user systems \cite{5425310, 8430512, 8955847, 7546944} or a Tx NULA with single antenna Rx users is assumed \cite{9049013, electronics6030057}.

In this paper, the \textit{joint} design of non-uniform Tx and Rx arrays of \textit{arbitrary} geometries for LoS MIMO systems is investigated towards maximizing the minimum capacity over a range of transmit distances, which also results in reduction of the capacity fluctuations. To formulate this as a closed-form optimization problem, we consider the capacities at a (finely quantized) set of transmit distances over the range. Then, by assuming a set of candidate antenna locations (depending on the assumed array geometry) for the Tx and Rx array, we express the joint array design problem as a multicriterion integer optimization problem (MIOP), which is NP-hard \cite{ehrgott2005multicriteria}. We demonstrate that, by relaxing the integer constraints, the joint optimization of the Tx and Rx arrays can be solved via convex optimization in a iterative manner \cite{1267055}. A local search method based on randomized rounding (RR) \cite{raghavan1987randomized} is also incorporated to improve the results. It is shown that the non-uniform configurations designed with our proposed approach outperform uniform as well as non-uniform array designs of the literature. To the best of the authors' knowledge, this is the first time that the LoS MIMO array design for optimizing the capacity is cast as a convex optimization problem.

The remainder of the paper is organized as follows. Section \ref{SecII} introduces the system model. The joint Tx-Rx array design problem is introduced as a MIOP in Section \ref{SecIII} and transformed into a convex optimization problem. Simulation results are presented in Section \ref{SecIV}, while Section \ref{SecV} concludes the paper. Bold lower case and upper case letters represent vectors and matrices, respectively. $(\cdot)^{\textit{T}}$ and $(\cdot)^\textit{H}$ denote the transpose and conjugate transpose of a vector or matrix.

\section{System Model}\label{SecII}
Consider a LoS MIMO system. The Tx and Rx arrays consist of $N$ and $M$ antennas, respectively and can admit arbitrary geometries and be arbitrarily oriented in space. The antenna locations of the Tx array are given by matrix $\textbf{T} \in \mathbb{R}^{3 \times N}$, where its $n-$th column is denoted as $\textbf{t}_{n}$ and represents the (x, y, z) Cartesian coordinates of the $n-$th Tx antenna on the 3D geometrical coordinate system, for $n \in \{1, ..., N\}$. The aperture of the array is given by the largest euclidean distance between any two elements of the array \cite{1543276} and is given by
\begin{equation}\label{eq:1}
	L_{\text{t}} =
	\underset{\begin{subarray}{c}
			i,j
	\end{subarray}}{\text{max}} \, \Vert \textbf{t}_{i} - \textbf{t}_{j} \Vert_{2}, \, \forall \; i, j \in \{1, \dots, N\}, \, i \neq j.
\end{equation}

The antenna locations of the Rx array are given by $\textbf{R} \in \mathbb{R}^{3 \times M}$, where the $m-$th column of $\textbf{R}$ is denoted as $\textbf{r}_{m}$ and represents the (x, y, z) Cartesian coordinates of the $m-$th Rx antenna on the 3D geometrical coordinate system, for $m \in \{1, ..., M\}$. The aperture of the Rx array $L_{\text{r}}$ is defined similarly to $L_{\text{t}}$. To avoid strong mutual coupling effects between the antennas, a minimum antenna spacing of $\lambda/2$ is considered at both arrays \cite{electronics6030057}, where $\lambda$ is the wavelength corresponding to carrier frequency $f_{\text{c}}$. Thus, we assume that 
\begin{subequations}\label{eq:2}
	\begin{align} 
		\underset{\begin{subarray}{c}
				i,j \\
		\end{subarray}}{\text{min}} \, \Vert \textbf{t}_{i} - \textbf{t}_{j} \Vert_{2} = d_{\text{t}} & \geq \lambda/2, \, \forall \; i, j \in \{1, \dots, N\}, \, i \neq j, \label{sub-eq-2:1} \\
		\underset{\begin{subarray}{c}
				i,j
		\end{subarray}}{\text{min}} \, \Vert \textbf{r}_{i} - \textbf{r}_{j} \Vert_{2} = d_{\text{r}} & \geq \lambda/2, \, \forall \; i, j \in \{1, \dots, M\}, \, i \neq j, \label{sub-eq-2:2}
	\end{align}
\end{subequations}
where $d_{\text{t}}$ and $d_{\text{r}}$ denote the minimum antenna spacing at the Tx and Rx array, respectively.

A LoS MIMO system is characterized by the expression
\begin{equation}\label{eq:3}
	\textbf{y} = \textbf{H x} + \textbf{n},
\end{equation}
which associates the received signal $\textbf{y} \in \mathbb{C}^{M}$ with the transmitted signal $\textbf{x} \in \mathbb{C}^{N}$, the channel matrix $\textbf{H} \in \mathbb{C}^{M \times N}$ and the complex additive white Gaussian noise of zero mean and unit variance, given by $\textbf{n} \in \mathbb{C}^{M}$. The power of $\textbf{x}$ is constrained to not be larger than $P_{\text{Tx}}$, i.e., $\Vert \textbf{x} \Vert_{2}^{2} \leq P_{\text{Tx}}$. The $(m, n)$-th entry of the LoS MIMO channel matrix is given as \cite{1543276, 7546944}
\begin{equation}\label{eq:4}
	\textbf{H}_{m, n} = e^{-j\frac{2\pi}{\lambda}d_{m, n}},
\end{equation}
where $d_{m, n} = \Vert \textbf{r}_{m} - \textbf{t}_{n} \Vert_{2}$ is the euclidean distance between the $m$-th Rx and $n$-th Tx antenna. In \eqref{eq:4} we assume that the transmit distance is much larger than the arrays’ apertures and that perfect power control is applied to compensate for the path loss, so that we can focus on the phase shifts of the channel matrix entries which are the determining factors of the spatial multiplexing capabilities of a MIMO system \cite{5425310, 8955847}.

By assuming no Tx channel state information, uniform power allocation across the Tx antennas is applied, hence, the capacity of the LoS MIMO system reads as \cite{1175470}
\begin{equation}\label{eq:5}
	\textit{C}  = \text{log}_2\left(\det \left(\textbf{I}_{M} + \frac{\rho}{N} \textbf{H} \textbf{H}^{\textit{H}}\right)\right), 
\end{equation}
where $\rho = \frac{P_{\text{Tx}}}{\sigma_{n}^2}$, with $\sigma_{n}^2$ being the noise variance at each receiving antenna and $\textbf{I}_{M}$ the $M \times M$ identity matrix. The maximum value of \eqref{eq:5} is achieved when all eigenvalues of $\textbf{H} \textbf{H}^{\textit{H}}$ are equal, so that the maximum capacity is given as
\begin{equation}\label{eq:6}
	C_{\text{max}} = M \, \text{log}_2 (1 + \rho).
\end{equation}

\section{Array Design}\label{SecIII}
The entries of the channel matrix \eqref{eq:4} depend only on the respective distance $d_{m, n}$ between the $n$-th Tx and $m$-th Rx antenna, with $n = 1, \dots, N$ and $m = 1, \dots, M$. Accordingly, these values depend on the transmit distance and on the locations of the Tx and Rx antennas, i.e., on matrices $\textbf{T}$ and $\textbf{R}$. Thus, the antennas' locations at the Tx and Rx array can be optimized in order to improve the capacity \cite{1175470, 1543276, 9363635, 5425310, 8430512, 9049013, 8955847, 7546944, electronics6030057}.

\subsection{Review of Uniform Array Design}\label{SecIIIa}
For a fixed transmit distance $D^{\ast}$, optimum uniform array configurations have been derived for LoS MIMO, so that maximum capacity is extracted at $D^{\ast}$. When ULAs are considered at the Tx and Rx, with an antenna spacing given by $d_{\text{t}}$ and $d_{\text{r}}$, it has been shown that \textit{maximum} capacity is extracted at $D^{\ast}$ if the following expression is satisfied \cite{1175470}
\begin{equation}\label{eq:7}
	d_{\text{t}} d_{\text{r}} = \frac{\lambda D^{\ast}}{\max(N, M)}.
\end{equation}
Similar expressions or optimization methods have been derived for UPAs \cite{1543276} and UCAs \cite{9363635}. Still, these results do not provide any insight about the capacity performance at other transmit distances. In fact, LoS MIMO systems suffer from substantial capacity fluctuations when operating at varying transmit distances when uniform arrays are employed \cite{5425310, 8430512}.

\begin{figure}[!t]
	\centering
	%	\large
	%\def\svgwidth{8.5cm}
	\def\svgwidth{0.9\columnwidth}
	\scalebox{1}{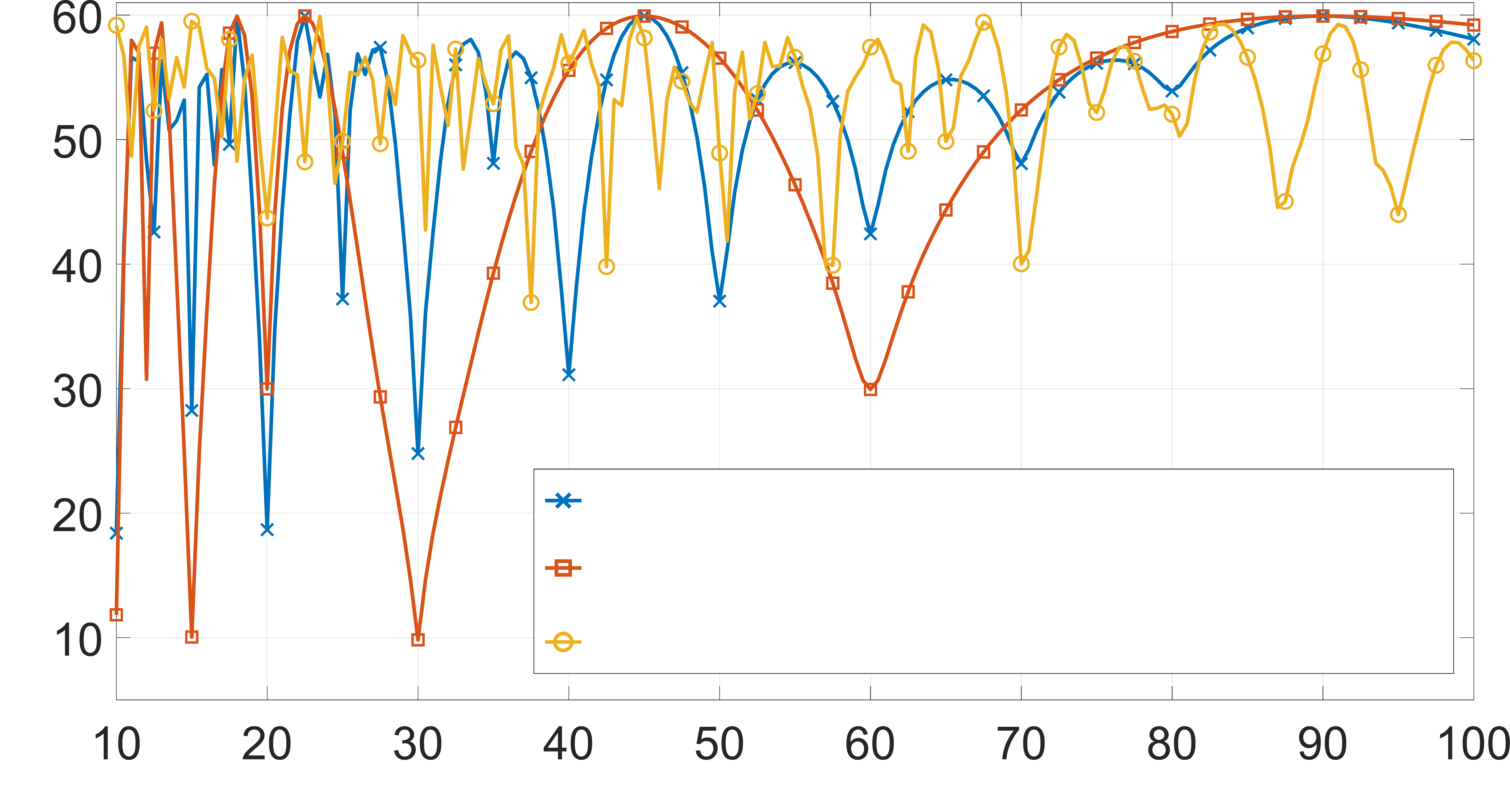}
	\caption{Capacity vs transmit distance for $9 \times 9$ uniform array topologies.}
	\label{fig:arrayComp}
\end{figure}

To showcase this limitation, the capacity performance of a LoS MIMO system over a range of distances is illustrated in Fig. \ref{fig:arrayComp}, where ULAs, UPAs and UCAs are assumed at the Tx and Rx. In all cases, $N = M = 9$. The ULAs, UPAs and UCAs were designed based on the results of \cite{1175470}, \cite{1543276} and \cite{9363635}, respectively, so that maximum capacity is extracted at the distance of $D^{\ast} = 90$ m. The carrier frequency is equal to $f_{\text{c}} = 62$ GHz. The Tx and Rx arrays have the same aperture size in each respective case and are equal to $1.76$ m, $1.08$ m and $3.12$ m for the ULA, UPA and UCA case, respectively. The capacity is evaluated over the range of $[10, 100]$ m.

To facilitate the evaluation and comparison of the capacity fluctuations, the mean $\mu$, standard deviation (std) $\sigma$ and minimum ($\text{min}$) of the capacity over the range of distances is depicted in the figure's legend, which are measured in bits-per-channel-use (bpcu). Evidently, all three LoS MIMO systems suffer from large capacity fluctuations over the given range. Although $N = M = 9$ is assumed here, this issue is also observed when different number of antennas is employed.

\subsection{Proposed Non-Uniform Array Design}\label{SecIIIb}
In order to improve the performance of LoS MIMO systems operating over a range of transmit distances, non-uniform design of the Tx and Rx arrays have been considered \cite{5425310, 8430512, 8955847, 9049013, 7546944, electronics6030057}. In this paper, we focus on the joint design of non-uniform Tx and Rx array configurations of arbitrary geometries for LoS MIMO systems. To the best of the authors' knowledge, \cite{7546944} is the only work that considered joint Tx and Rx non-uniform array design, albeit, only NULAs were considered.

First, the objective function, that is to be maximized, should be the capacity over a range of transmit distances $\textbf{d}$. To express this optimization problem in a closed form, we introduce a \textit{quantization} of the range. In particular, we consider the range, not as a continuum, but as a discrete set of transmit distances, that is, we suggest that $\textbf{d}$ is represented as a $Q \times 1$ vector $\textbf{d} = [D_{1}, \dots, D_{q}, \dots, D_{Q}]^{\textit{T}}$, with $D_{1}$ and $D_{Q}$ being the minimum and maximum distances of the range, and $Q$ the number of considered distances. Note that optimizing the quantization step (i.e., $Q$) is out of the scope of this paper, however, it has been shown, by means of simulations, that the capacity of a LoS MIMO system is robust to small changes in the transmit distance \cite{1543276, 8356240}. So, we would expect that a sufficiently small value of $Q$ would be adequate for a proper evaluation of the capacity over the range of distances.

So, the capacity at transmit distance $D_{q}$ can be written as
\begin{equation}\label{eq:8}
	\textit{C}_{q}(\textbf{T}, \textbf{R})  = \text{log}_2\left(\det \left(\textbf{I}_{M} + \frac{\rho}{N} \textbf{H}^{q} \textbf{H}^{q^{\textit{H}}}\right)\right), 
\end{equation}
where $\textbf{H}^{q} \in \mathbb{C}^{M \times N}$ is the channel matrix realization at the transmit distance $D_{q}$. The notation in \eqref{eq:8} highlights the dependence of the capacity on the array configurations, which originates from the dependence of $\textbf{H}^{q}$ on $\textbf{T}$ and $\textbf{R}$ (see \eqref{eq:4}).

To simplify the array design problem, we focus on optimally selecting antenna locations, for a given Tx and Rx array geometry, from a \textit{uniform grid of candidate antenna locations}, so that the array design problem becomes an antenna location selection problem. So, consider that the $N$ Tx and $M$ Rx antennas can be placed only on $N_{\text{f}}$ and $M_{\text{f}}$ possible locations, respectively, which are uniformly spaced across the arrays' physical areas, with $N_{\text{f}} \geq N$ and $M_{\text{f}} \geq M$. We highlight that our approach can support any array geometry type. Let the coordinates of the $N_{\text{f}}$ candidate locations of the Tx antennas be given by the columns of matrix $\textbf{T}_{\text{f}} \in \mathbb{R}^{3 \times N_{\text{f}}}$, which obey
\begin{equation}\label{eq:9}
		\underset{\begin{subarray}{c}
		i,j \\
\end{subarray}}{\text{max}} \, \Vert \textbf{t}_{i} - \textbf{t}_{j} \Vert \leq L_{\text{t}}, \, \forall \; i, j \in \{1, \dots, N_{\text{f}}\}, \, i \neq j
\end{equation}
as well as \eqref{sub-eq-2:1}. Similarly, the coordinates of the $M_{\text{f}}$ candidate locations of the Rx antennas are given by the columns of matrix $\textbf{R}_{\text{f}} \in \mathbb{R}^{3 \times M_{\text{f}}}$, which must satisfy a constraint similar to \eqref{eq:9} as well as \eqref{sub-eq-2:2}. Note that the grid of antenna locations can be non-uniform too, as long as \eqref{sub-eq-2:1}, \eqref{sub-eq-2:2} and \eqref{eq:9} hold.

Let $\textbf{H}_{\text{f}}^{q} \in \mathbb{C}^{M_{\text{f}} \times N_{\text{f}}}$ be the channel matrix at transmit distance $D_{q}$ between the Tx and Rx arrays, assuming that antennas are placed on \textit{all} candidate locations. The entries of $\textbf{H}_{\text{f}}^{q}$ are defined as in \eqref{eq:4}. Given that each row of the MIMO channel matrix corresponds to a specific Rx antenna and that each column corresponds to a specific Tx antenna, performing antenna (location) selection on $\textbf{R}_{\text{f}}$ is equivalent to performing \textit{row} selection on $\textbf{H}_{\text{f}}^{q}$, while performing antenna (location) selection on $\textbf{T}_{\text{f}}$ is equivalent to performing \textit{column} selection on $\textbf{H}_{\text{f}}^{q}$. 

As our goal is to select $N$ out of $N_{\text{f}}$ Tx and $M$ out of $M_{\text{f}}$ Rx antenna locations, if the coordinates of the Tx and Rx \textit{selected} antennas are given by $\textbf{T} \in \mathbb{R}^{3 \times N}$ and $\textbf{R} \in \mathbb{R}^{3 \times M}$, respectively, then by introducing Tx and Rx \textit{binary selection} matrices as $\textbf{F}^{\text{t}} \in \mathbb{B}^{N_{\text{f}} \times N}$ and $\textbf{F}^{\text{r}} \in \mathbb{B}^{M_{\text{f}} \times M}$, we have that $\textbf{T} = \textbf{T}_{\text{f}} \textbf{F}^{\text{t}}$ and $\textbf{R} = \textbf{R}_{\text{f}} \textbf{F}^{\text{r}}$. Matrices $\textbf{F}^{\text{t}}$ and $\textbf{F}^{\text{r}}$ have a single one in each column and at most a single one in each row \cite{7370753}. To that end, the channel matrix between the Tx and Rx arrays, whose antenna locations have been optimally selected, is given as $\textbf{H}^{q} = \textbf{F}^{\text{r}^{\textit{H}}} \textbf{H}^{q}_{\text{f}} \textbf{F}^{\text{t}}$. So, \eqref{eq:8} can be now written as a function of $\textbf{F}^{\text{t}}$ and $\textbf{F}^{\text{r}}$ as
\begin{align}\label{eq:10}
		\textit{C}_{q}(\textbf{F}^{\text{t}}, \textbf{F}^{\text{r}}) &= \text{log}_2\left(\det \left(\textbf{I}_{M} + \frac{\rho}{N_{\text{f}}} (\textbf{F}^{\text{r}^{\textit{T}}} \textbf{H}^{q}_{\text{f}} \textbf{F}^{\text{t}}) (\textbf{F}^{\text{r}^{\textit{T}}} \textbf{H}^{q}_{\text{f}} \textbf{F}^{\text{t}})^{\textit{H}}\right)\right) \notag \\	
		\Rightarrow \textit{C}_{q}(\boldsymbol{\Delta}^{\text{t}}, \boldsymbol{\Delta}^{\text{r}}) &= \text{log}_2\left(\det \left(\textbf{I}_{M} + \frac{\rho}{N_{\text{f}}} \textbf{H}_{\text{f}}^{q} \boldsymbol{\Delta}^{\text{t}} \textbf{H}_{\text{f}}^{q^{\textit{H}}} \boldsymbol{\Delta}^{\text{r}} \right)\right),
\end{align}
where $\boldsymbol{\Delta}^{\text{t}} = \textbf{F}^{\text{t}} \textbf{F}^{\text{t}^{\textit{T}}}$ and $\boldsymbol{\Delta}^{\text{r}} = \textbf{F}^{\text{r}} \textbf{F}^{\text{r}^{\textit{T}}}$ and we have applied Sylvester's determinant identity in \eqref{eq:10} \cite{boyd2004convex}. $\boldsymbol{\Delta}^{\text{t}} \in \mathbb{R}^{N_{\text{f}} \times N_{\text{f}}}$ and $\boldsymbol{\Delta}^{\text{r}} \in \mathbb{R}^{M_{\text{f}} \times M_{\text{f}}}$ are binary diagonal matrices that act as Tx and Rx selection matrices, respectively.

In particular, $\Delta^{\text{t}}_{n_{\text{f}}} = 1$ (for notational brevity, $\Delta_{i}$ denotes the $i$-th diagonal element of diagonal matrix $\boldsymbol{\Delta}$), where $n_{\text{f}} \in \{1, \dots, N_{\text{f}}\}$ indicates possible antenna locations of the Tx array, signifies that the $n_{\text{f}}$-th candidate antenna location is selected at the Tx array, while $\Delta^{\text{t}}_{n_{\text{f}}} = 0$ indicates otherwise. $\Delta^{\text{r}}_{m_{\text{f}}}$ is defined similarly for the Rx array, where $m_{\text{f}} \in \{1, \dots, M_{\text{f}}\}$, with $\Delta^{\text{r}}_{m_{\text{f}}} = 1$ indicating that the $m_{\text{f}}$-th candidate antenna location is selected at the Rx array. As $N$ out of $N_{\text{f}}$ Tx antennas and $M$ out of $M_{\text{f}}$ Rx antennas must be selected, the number of ones in the diagonal of $\boldsymbol{\Delta}^{\text{t}}$ and $\boldsymbol{\Delta}^{\text{r}}$ is equal to $N$ and $M$, respectively. So, the joint Tx and Rx array design problem is formulated as an antenna (location) selection problem as
\begin{subequations}
	\begin{align} 
	\underset{\boldsymbol{\Delta}^{\text{t}}, \, \boldsymbol{\Delta}^{\text{r}}} {\text{maximize}} \; &\ \{C_{1}(\boldsymbol{\Delta}^{\text{t}}, \boldsymbol{\Delta}^{\text{r}}), \dots, C_{Q}(\boldsymbol{\Delta}^{\text{t}}, \boldsymbol{\Delta}^{\text{r}})\}, \label{eq:12a} \\
	\text{s.t.:} &\ \; \sum_{n_{\text{f}} = 1}^{N_{\text{f}}} \Delta_{n_{\text{f}}}^{\text{t}} = N, \label{eq:12b}  \\
	&\ \; \sum_{m_{\text{f}} = 1}^{M_{\text{f}}} \Delta_{m_{\text{f}}}^{\text{r}} = M, \label{eq:12c}  \\
	&\ \; \Delta_{n_{\text{f}}}^{\text{t}} \in \{0, 1\}, \, \forall \, n_{\text{f}} \in \{1, \dots, N_{\text{f}}\}, \label{eq:12d} \\
	&\ \; \Delta_{m_{\text{f}}}^{\text{r}} \in \{0, 1\}, \, \forall \, m_{\text{f}} \in \{1, \dots, M_{\text{f}}\}. \label{eq:12e} 
	\end{align}
\end{subequations}

\eqref{eq:12a} - \eqref{eq:12e} is a MIOP which is NP-hard \cite{ehrgott2005multicriteria}. Solving it amounts to identifying all Pareto optimal points \cite{boyd2004convex}. As each such point would correspond to a different Tx and Rx non-uniform array configuration, we do not search for the Pareto front of \eqref{eq:12a} - \eqref{eq:12e}, but focus instead on maximizing the \textit{minimum} capacity across the range of transmit distances. In this way, we can obtain Tx and Rx non-uniform array configurations that reduce the capacity fluctuations across the range. Thus, we consider the following optimization
\begin{subequations}
	\begin{align} 
	\underset{\boldsymbol{\Delta}^{\text{t}}, \, \boldsymbol{\Delta}^{\text{r}}} {\text{maximize}} \; &\ \underset{q = 1, \dots, Q} {\text{min}} \{C_{1}(\boldsymbol{\Delta}^{\text{t}}, \boldsymbol{\Delta}^{\text{r}}), \dots, C_{Q}(\boldsymbol{\Delta}^{\text{t}}, \boldsymbol{\Delta}^{\text{r}})\}, \label{eq:13a} \\
	\text{s.t.:} &\ \; \sum_{n_{\text{f}} = 1}^{N_{\text{f}}} \Delta_{n_{\text{f}}}^{\text{t}} = N, \label{eq:13b}  \\
	&\ \; \sum_{m_{\text{f}} = 1}^{M_{\text{f}}} \Delta_{m_{\text{f}}}^{\text{r}} = M, \label{eq:13c}  \\
	&\ \; \Delta_{n_{\text{f}}}^{\text{t}} \in \{0, 1\}, \, \forall \, n_{\text{f}} \in \{1, \dots, N_{\text{f}}\}, \label{eq:13d} \\
	&\ \; \Delta_{m_{\text{f}}}^{\text{r}} \in \{0, 1\}, \, \forall \, m_{\text{f}} \in \{1, \dots, M_{\text{f}}\}. \label{eq:13e} 
	\end{align}
\end{subequations}
An optimum solution of \eqref{eq:13a} - \eqref{eq:13e} can be found via an exhaustive search (ES) over all antenna locations and all $Q$ transmit distances. However, as this entails a complexity of $\Ob(\binom{N_{\text{f}}}{N} \binom{M_{\text{f}}}{M} Q)$ operations, it is only applicable for very small values of $N_{\text{f}}, M_{\text{f}}, N, M$ and $Q$. In fact, prior works that adopt ES make the restrictive assumptions that the Tx and Rx arrays have identical configurations and that $N = M \leq 4$ \cite{5425310, 8430512}. Instead, we allow the Tx and Rx arrays to have different configurations and we cast \eqref{eq:13a} - \eqref{eq:13e} as a convex optimization problem, which can be efficiently solved.

First, the integer constraints \eqref{eq:13d} and \eqref{eq:13e} can be relaxed by assuming that the diagonal entries of $\boldsymbol{\Delta}^{\text{t}}$ and $\boldsymbol{\Delta}^{\text{r}}$ can take any real value between zero and one \cite{1687757, 4663892, 9229166}, so that \eqref{eq:13a} - \eqref{eq:13e} is written as
\begin{subequations}
	\begin{align} 
	\underset{\boldsymbol{\Delta}^{\text{t}}, \, \boldsymbol{\Delta}^{\text{r}}} {\text{maximize}} \; &\ \underset{q = 1, \dots, Q} {\text{min}} \{C_{1}(\boldsymbol{\Delta}^{\text{t}}, \boldsymbol{\Delta}^{\text{r}}), \dots, C_{Q}(\boldsymbol{\Delta}^{\text{t}}, \boldsymbol{\Delta}^{\text{r}})\}, \label{eq:14a} \\
	\text{s.t.:} &\ \; \sum_{n_{\text{f}} = 1}^{N_{\text{f}}} \Delta_{n_{\text{f}}}^{\text{t}} = N, \label{eq:14b}  \\
	&\ \; \sum_{m_{\text{f}} = 1}^{M_{\text{f}}} \Delta_{m_{\text{f}}}^{\text{r}} = M, \label{eq:14c}  \\
	&\ \; 0 \leq \Delta_{n_{\text{f}}}^{\text{t}} \leq 1, \, \forall \, n_{\text{f}} \in \{1, \dots, N_{\text{f}}\}, \label{eq:14d} \\
	&\ \; 0 \leq \Delta_{m_{\text{f}}}^{\text{r}} \leq 1, \, \forall \, m_{\text{f}} \in \{1, \dots, M_{\text{f}}\}. \label{eq:14e} 
	\end{align}
\end{subequations}
Secondly, we note that a function $f(\textbf{A}) = \log_{2} \det (\textbf{A})$ is concave in the elements of $\textbf{A}$ if $\textbf{A}$ is positive definite and that concavity is preserved under affine transformations \cite{1687757, boyd2004convex}. Thirdly, we point out that the pointwise minimum of a set of concave functions is also a concave function \cite{boyd2004convex}. Given the second point, regarding the $q$-th transmit distance, $C_{q}(\boldsymbol{\Delta}^{\text{t}}, \boldsymbol{\Delta}^{\text{r}})$ in \eqref{eq:10} is concave in the elements of $\boldsymbol{\Delta}^{\text{t}}$ or $\boldsymbol{\Delta}^{\text{r}}$, but it is not jointly concave in the elements of both $\boldsymbol{\Delta}^{\text{t}}$ and $\boldsymbol{\Delta}^{\text{r}}$, since the product of positive semidefinite matrices is not jointly concave in the elements of the matrices \cite{boyd2004convex}. However, If \eqref{eq:14a} - \eqref{eq:14e} is solved \textit{iteratively}, i.e., by fixing one of the optimization variables at each iteration and finding the optimum solution for the other, convex optimization techniques can be employed \cite{1267055, 9229166}. The procedure is repeated until the objective function does not improve beyond a given threshold.

Thus, we propose to solve \eqref{eq:14a} - \eqref{eq:14e} in an iterative fashion, as it is concave in $\boldsymbol{\Delta}^{\text{t}}$ when $\boldsymbol{\Delta}^{\text{r}}$ is fixed and vice versa. At each iteration, the associated problem can be efficiently solved in polynomial time via interior-point methods. By employing the barrier method \cite{boyd2004convex}, the computational complexity of the problem at each iteration is dictated by the Cholesky factorization at each Newton step, which is equal to $\Ob(N_{\text{f}}^{3})$ and $\Ob(M_{\text{f}}^{3})$ for the optimization of $\boldsymbol{\Delta}^{\text{t}}$ and $\boldsymbol{\Delta}^{\text{r}}$, respectively. The number of Newton steps is upper bounded by $\Ob(\sqrt{N_{\text{f}}})$ and $\Ob(\sqrt{M_{\text{f}}})$ \cite{boyd2004convex}. If $P$ is the number of iterations that the iterative maximization algorithm takes to converge, then the complexity of solving \eqref{eq:14a} - \eqref{eq:14e} is $\Ob(P Q (N_{\text{f}}^{3.5} + M_{\text{f}}^{3.5}))$, which is significantly lower than the ES complexity.

After the iterative algorithm terminates, the optimum $\boldsymbol{\Delta}^{\text{t}^{\ast}}$ and $\boldsymbol{\Delta}^{\text{r}^{\ast}}$ are obtained. However, due to the relaxation, the diagonal elements of these matrices can be fractional values, so they would not constitute a feasible solution for the initial problem \eqref{eq:13a}. To overcome this, we may consider the indices of the $N$ and $M$ \textit{largest} diagonal elements of $\boldsymbol{\Delta}^{\text{t}^{\ast}}$ and $\boldsymbol{\Delta}^{\text{r}^{\ast}}$, respectively, as the optimal antenna positions of the Tx and Rx arrays \cite{1687757}. 

To refine the initial solution, a more sophisticated selection scheme can be considered. For instance, randomized rounding (RR) is a well-known approximation algorithm in combinatorial optimization \cite{raghavan1987randomized}, which is of low computational complexity and can be incorporated into problems with relaxation of integer constraints \cite{4663892, 9229166}. RR can be applied to refine the solutions $\boldsymbol{\Delta}^{\text{t}^{\ast}}$ and $\boldsymbol{\Delta}^{\text{r}^{\ast}}$. Considering $\boldsymbol{\Delta}^{\text{t}^{\ast}}$, the aim is to group its $N_{\text{f}}$ diagonal elements into two sets: a first set with $N$ elements whose indices indicate the selected antenna locations of the Tx array, and a second set containing the remaining $N_{\text{f}} - N$ elements. Initially, the $N$ element with the largest magnitude in the first set are sorted in a descending order.

RR allows elements from the first set to be swapped by elements of the second set, thereby refining the initial $\boldsymbol{\Delta}^{\text{t}^{\ast}}$. Specifically, we go through each element of the first set in an ascending order (smallest to largest) and replace it with an element of the second set (going through all elements of this set), if the value of the objective function can be increased with this swapping. The algorithm terminates when all possible swaps have been checked, which amounts to a complexity of $\Ob(N(N_{\text{f}} - N))$ or when the value of the objective function is above a predetermined threshold. An identical procedure can be applied for refining the solution of $\boldsymbol{\Delta}^{\text{r}^{\ast}}$ too. By doing this, we can obtain antenna locations for the Tx and Rx arrays that can achieve a larger minimum capacity compared to the initial solution $\boldsymbol{\Delta}^{\text{t}^{\ast}}$ and $\boldsymbol{\Delta}^{\text{r}^{\ast}}$ of \eqref{eq:14a}. As the complexity of this local search is equal to $\Ob(N(N_{\text{f}} - N) + M(M_{\text{f}} - M))$, it is negligible compared to the complexity of solving \eqref{eq:14a} - \eqref{eq:14e}.

\section{Numerical Results}\label{SecIV}
In this section, the performance of non-uniform array configurations designed via the proposed convex optimization (CO) approach (with and without RR) is evaluated over a range of transmit distances. Due to space limitations, we focus on linear arrays, but point out that our approach is applicable to any array geometry. We compare the performance of the proposed NULAs against that of ULAs and of NULA designs from the literature. Without loss of generality, the Tx and Rx arrays are assumed to be aligned and facing each other. We assume $f_{\text{c}}$ = 62 GHz, $\rho = 20$ dB and that the transmit distance lies within the range of $[10, 100]$ m. We consider a quantization step of 0.5 m for the given range, which translates to a transmit distance vector $\textbf{d}$ consisting of $Q = 181$ elements.

For our comparisons, we assume ULAs at the Tx and Rx of equal number of antennas which are optimized at $D^{\ast} = 92$ m (considering smaller $D^{\ast}$ led to poorer performance within the given range for the ULAs). For $M = N = 4$, we have $L_{\text{t}} = L_{\text{r}} = 1$ m from \eqref{eq:7}. So, the uniform antenna spacing at both arrays is equal to $d_{\text{t}} = d_{\text{r}} = L_{\text{t}} / (N -1) = 0.33$ m. For the NULAs design, we assume the same aperture as the Tx and Rx ULAs and that there are $N_{\text{f}} = M_{\text{f}} = 16$ candidate antenna locations, uniformly spaced within the assumed apertures, resulting in a spacing of $d_{\text{t,f}} = d_{\text{r,f}} = L_{\text{t}} / (N_{\text{f}} - 1) = 0.067$ m.

\begin{figure}[!t]
	\centering
	\def\svgwidth{0.9\columnwidth}
	\scalebox{1}{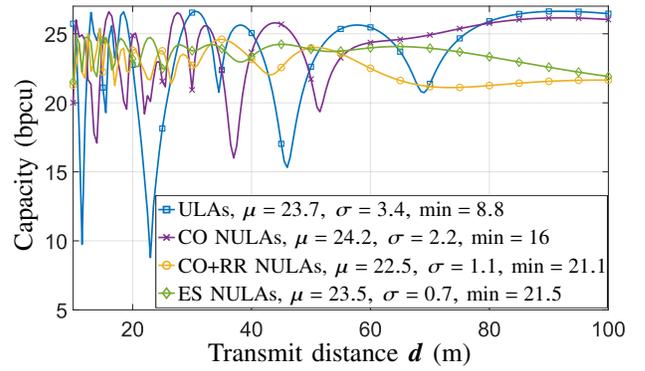}
	\caption{Capacity versus (vs.) transmit distance of proposed NULAs and ULAs.}
	\label{fig:NULAs16_cap}
\end{figure}

The NULAs' performance designed with the iterative CO approach is illustrated in Fig. \ref{fig:NULAs16_cap}. The iterations terminate when the increase in the objective function in \eqref{eq:14a} becomes less than a threshold of $0.01$ bpcu between consecutive iterations, which results in $P = 5$ iterations. To highlight the effect of the local search, results with and without RR are illustrated in Fig. \ref{fig:NULAs16_cap}. This figure also includes the performance achieved with ULAs and with the optimum NULAs obtained via an ES over all $\binom{N_{\text{f}}}{N} \binom{M_{\text{f}}}{M}$ possible combinations of the antenna locations. For each configuration, we provide in the legend the mean, standard deviation and minimum of the capacities achieved within the range of distances. Our proposed NULA designs, with and without RR, distinctly outperform the ULAs in terms of minimum capacity over the given range of transmit distances. We also notice that RR can greatly enhance the performance, since the CO+RR NULAs lead to a much higher minimum capacity compared to the CO NULAs. Remarkably, the performance of the NULAs designed with the CO+RR approach is very similar to the one of the ES-based NULAs.

\begin{figure}[!t]
	\centering
	\def\svgwidth{0.9\columnwidth}
	\subfloat[]{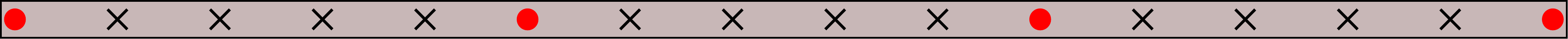
		\label{fig:TxbaselineULA16}}
		\hfil
	\def\svgwidth{0.9\columnwidth}
	\subfloat[]{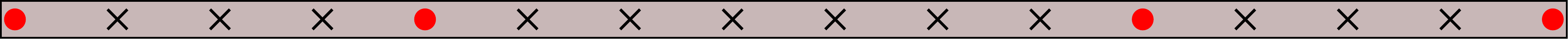
		\label{fig:TxRxCO_NULA16}}
	\hfil
	\def\svgwidth{0.9\columnwidth}
	\subfloat[]{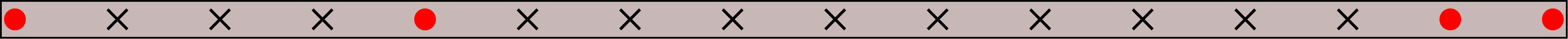
		\label{fig:TxCORR_NULA16}}
	\hfil
	\def\svgwidth{0.9\columnwidth}
	\subfloat[]{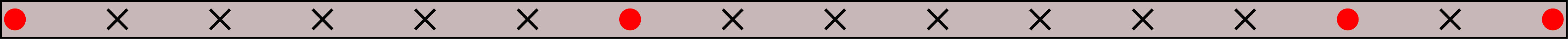
		\label{fig:RxCORR_NULA16}}
		\hfil
	\def\svgwidth{0.9\columnwidth}
	\subfloat[]{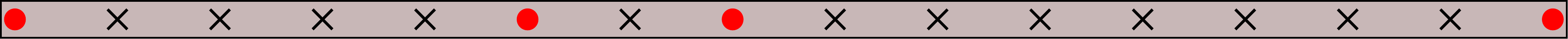
		\label{fig:TxES_NULA16}}
		\hfil
	\def\svgwidth{0.9\columnwidth}
	\subfloat[]{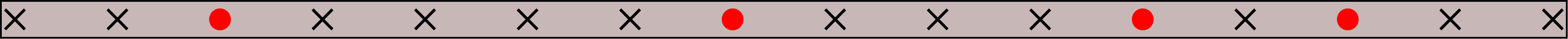
		\label{fig:RxES_NULA16}}
	\caption{(a) Tx/Rx ULA, (b) Tx/Rx CO NULA, (c) Tx CO+RR NULA, (d) Rx CO+RR NULA, (e) Tx ES NULA and (f) Rx ES NULA. Crosses and circles indicate candidate and selected antenna locations, respectively.}
	\label{fig:NULAs16}
\end{figure}

The arrays used in Fig. \ref{fig:NULAs16_cap} are shown in Fig. \ref{fig:NULAs16}. Evidently, the Tx and Rx admit the same NULA geometry when the CO approach without RR is used. Although all NULAs designed with our proposed methods have the same aperture size as the ULAs in this particular scenario, in general, this is not a prerequisite for our approach, since the designed arrays can have, in principle, a smaller aperture than the initially assumed aperture. This is an important advantage of our scheme, since, in the literature, a fixed aperture size is usually assumed, i.e., that the locations of the edge antennas are \textit{fixed}, and only the positions of inner elements are optimized \cite{5425310, 8430512}.

\begin{figure}[!t]
	\centering
	\def\svgwidth{0.9\columnwidth}
	\scalebox{1}{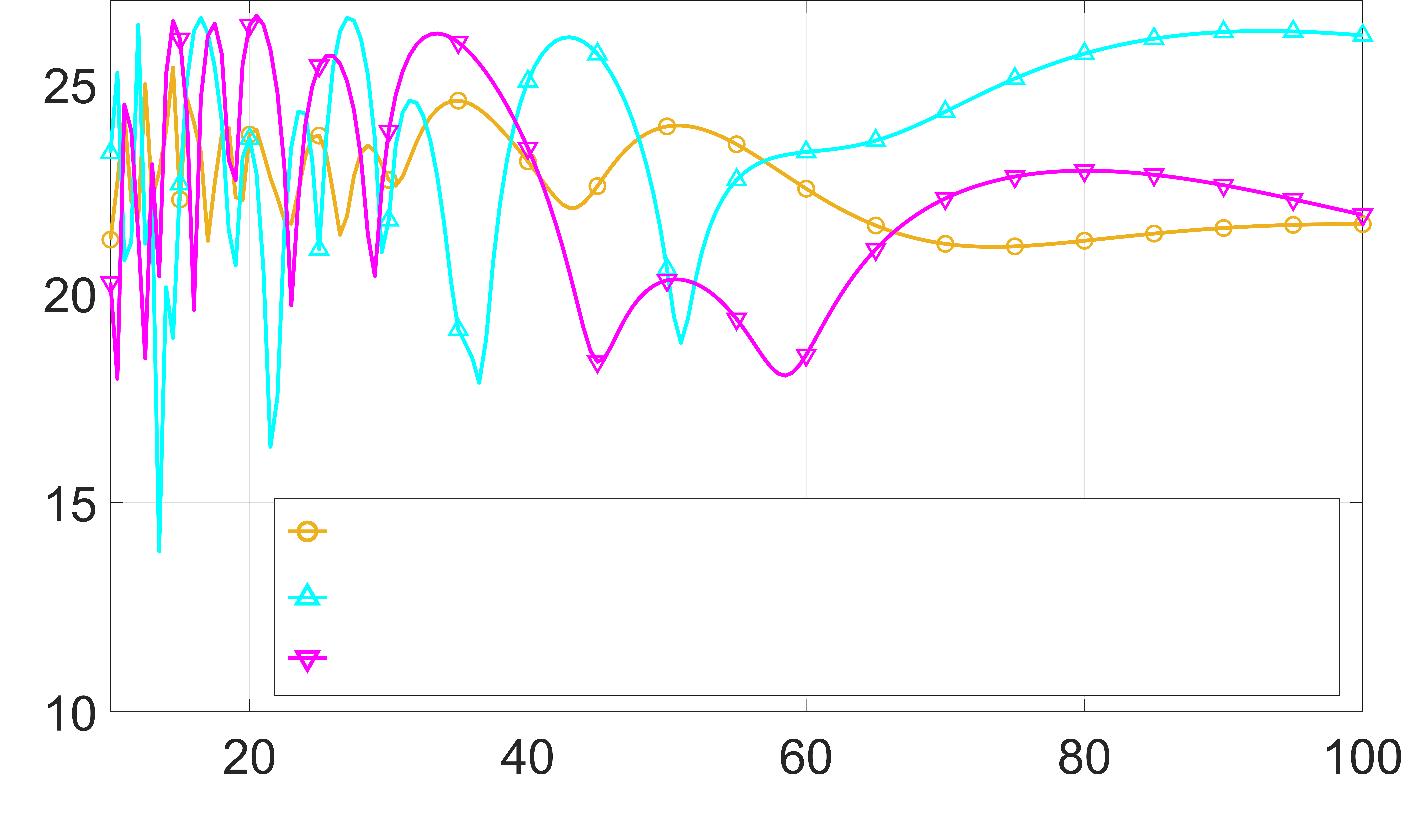}
	\caption{Capacity vs. transmit distance of proposed and state-of-the-art NULAs.}
	\label{fig:NULAs16_cap_comp} 
\end{figure}

Regarding NULA designs of the literature, by considering the ES design in Fig. \ref{fig:NULAs16_cap}, we provide a comparison of our proposed approaches with an improved version of the schemes presented in \cite{5425310, 8430512}, as the ES in \cite{5425310, 8430512} assumes the same NULA configuration at the Tx and Rx. In Fig. \ref{fig:NULAs16_cap_comp}, we compare the NULAs derived via the CO+RR approach with the ones in \cite{7546944} and \cite{electronics6030057}. Although the proposal in \cite{electronics6030057} is associated with a system where only the Tx array (surrounded by single antenna Rx users) is optimized, in Fig. \ref{fig:NULAs16_cap_comp} we assume the optimally derived array at both the Tx and Rx (this leads to better performance, compared to assuming a ULA at the Rx). The complexity of \cite{7546944} is $\Ob(\max(N^3, M^3))$, due to a QR decomposition of the matrices whose elements correspond to the antenna positions of the Tx and Rx. The complexity of \cite{electronics6030057} is equal to $\Ob(A Q)$, where $A$ is the number of possible values of a parameter $\alpha_1$ which characterizes the NULA parameterization \cite{electronics6030057}. In Fig. \ref{fig:NULAs16_cap_comp}, $\alpha_1 = 0.205$ is used, as this lead to best performance. Our proposed designs outperform the ones in \cite{7546944, electronics6030057} in terms of minimum capacity, standard deviation and, thus, of capacity fluctuations over the considered range.

\begin{figure}[!t]
	\centering
	\def\svgwidth{0.87\columnwidth}
	\scalebox{1}{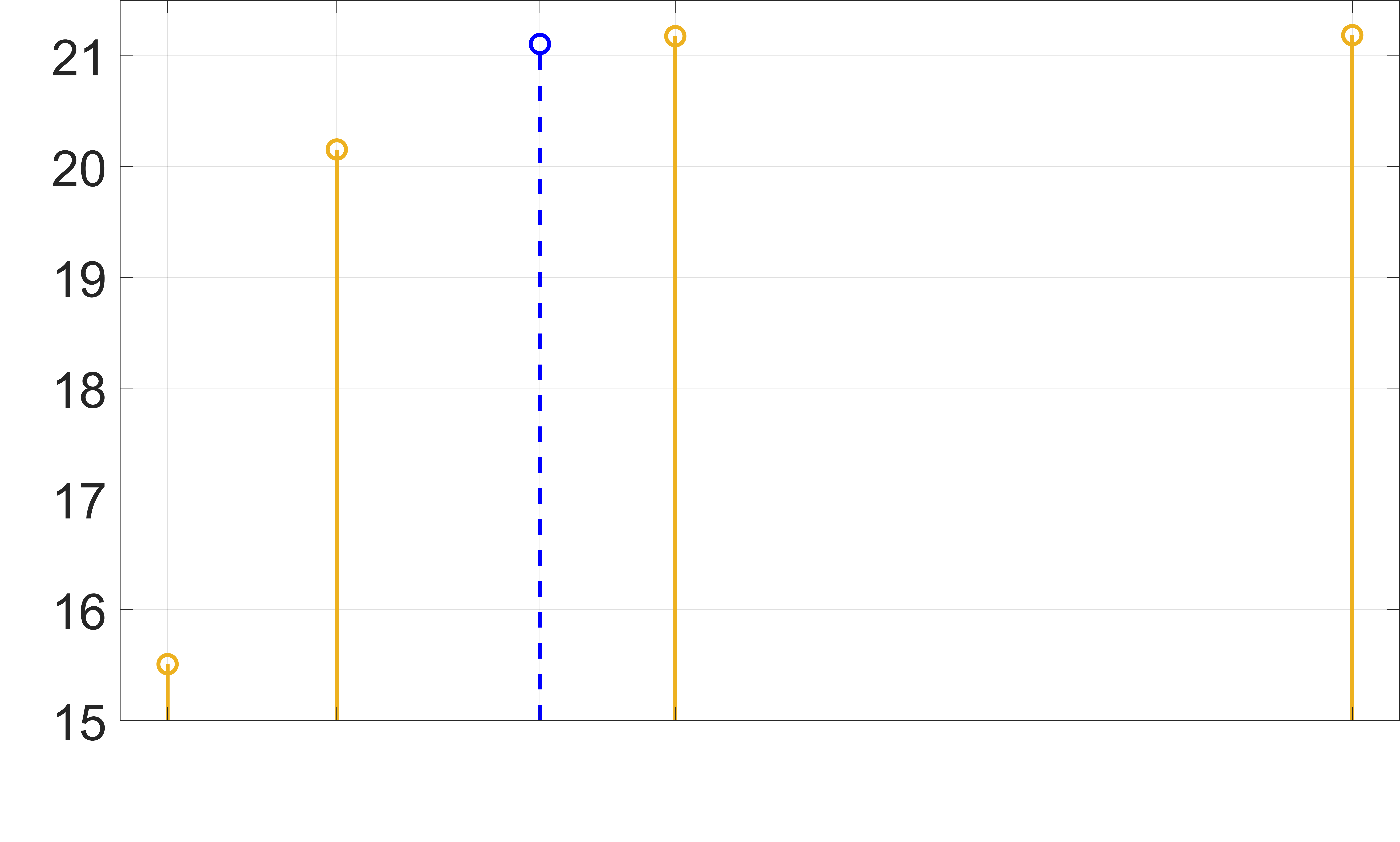}
	\caption{Min. capacity vs. different Tx and Rx uniform grid granularities.}
	\label{fig:NULAs_multiNfMf}
\end{figure}

The min. capacity vs. the spacing $d_{\text{t,f}}$ of the (uniform) grid of candidate antenna locations is shown in Fig. \ref{fig:NULAs_multiNfMf} for the CO+RR approach, where $N = M = 4$ and the same Tx and Rx grid spacing ($d_{\text{t,f}} = d_{\text{r,f}}$) is assumed. We assume increasingly smaller spacing, so that the effect of the grid's granularity can be evaluated. For a grid spacing $d_{\text{t,f}}$ that is $\eta$ times finer than the ULA spacing, i.e., $d_{\text{t,f}} = d_{\text{t}}/\eta$, the number of candidate antenna locations is given as $N_{\text{f}} = \eta(N - 1) + 1$. For example, for $d_{\text{t,f}} = d_{\text{t}}/8$, we have that $N_{\text{f}} = 25$. The same holds for the Rx grid. As expected, the min. capacity is increased for grids of finer granularity, that is, of higher number of candidate antenna locations. However, it seems that only marginal improvement is attained when the spacing becomes too small. The minimum allowable grid spacing is dictated by \eqref{sub-eq-2:1} and \eqref{sub-eq-2:2} (both are satisfied in Fig. \ref{fig:NULAs_multiNfMf}). For comparison, $d_{\text{t,f}} = d_{\text{r,f}} = d_{\text{t}}/5$ which corresponds to $N_{\text{f}} = M_{\text{f}} = 16$ that was used in Fig. \ref{fig:NULAs16_cap} - \ref{fig:NULAs16_cap_comp} is also illustrated in Fig. \ref{fig:NULAs_multiNfMf} with a dotted line.

\section{Conclusions}\label{SecV}
We proposed a novel method for the joint design of non-uniform Tx and Rx arrays of arbitrary geometries for LoS MIMO systems towards maximizing the minimum capacity over a range of transmit distances. The method was based on optimally selecting antenna locations from a grid of candidate antenna locations. We demonstrated, by leveraging convex relaxation, that the joint Tx and Rx array design problem can be formulated as a convex optimization problem, solved in polynomial time, if it is solved iteratively. A local optimization method based on RR was also incorporated to improve the performance. The non-uniform arrays designed with our approach featured superior capacity performance compared to uniform and non-uniform array configurations of the literature and comparable performance to the ES-based designs, despite the much lower computational complexity of our approach.

%\ifCLASSOPTIONcaptionsoff
%\newpage
%\fi

%\bibliographystyle{myIEEEtran}
\bibliographystyle{IEEEtran}
\bibliography{IEEEabrv, paper_draft}

\end{document}

%% file: figures/arraysComp_v5.pdf_tex
%% Creator: Inkscape 1.2.1 (9c6d41e410, 2022-07-14), www.inkscape.org
%% Accompanies image file 'arraysComp_v5.pdf' (pdf, eps, ps)
\begingroup%
  \makeatletter%
  \providecommand\color[2][]{%
    \errmessage{(Inkscape) Color is used for the text in Inkscape, but the package 'color.sty' is not loaded}%
    \renewcommand\color[2][]{}%
  }%
  \providecommand\transparent[1]{%
    \errmessage{(Inkscape) Transparency is used (non-zero) for the text in Inkscape, but the package 'transparent.sty' is not loaded}%
    \renewcommand\transparent[1]{}%
  }%
  \providecommand\rotatebox[2]{#2}%
  \newcommand*\fsize{\dimexpr\f@size pt\relax}%
  \newcommand*\lineheight[1]{\fontsize{\fsize}{#1\fsize}\selectfont}%
  \ifx\svgwidth\undefined%
    \setlength{\unitlength}{2482.57635498bp}%
    \ifx\svgscale\undefined%
      \relax%
    \else%
      \setlength{\unitlength}{\unitlength * \real{\svgscale}}%
    \fi%
  \else%
    \setlength{\unitlength}{\svgwidth}%
  \fi%
  \global\let\svgwidth\undefined%
  \global\let\svgscale\undefined%
  \makeatother%
  \begin{picture}(1,0.5379287)%
    \lineheight{1}%
    \setlength\tabcolsep{0pt}%
    \put(0,0){\includegraphics[width=\unitlength,page=1]{arraysComp_v5.pdf}}%
    \put(0.39151759,0.194){\color[rgb]{0,0,0}\makebox(0,0)[lt]{\lineheight{1.25}\smash{\footnotesize{\begin{tabular}[t]{l}ULAs, $\mu = 52.7$, $\sigma = 7.6$, $\text{min} = 18.4$ \end{tabular}}}}}%
    \put(0.39151759,0.1491){\color[rgb]{0,0,0}\makebox(0,0)[lt]{\lineheight{1.25}\smash{\footnotesize{\begin{tabular}[t]{l}UPAs, $\mu = 49.4$, $\sigma = 12.5$, $\text{min} = 9.8$ \end{tabular}}}}}%
    \put(0.39151759,0.10){\color[rgb]{0,0,0}\makebox(0,0)[lt]{\lineheight{1.25}\smash{\footnotesize{\begin{tabular}[t]{l}UCAs, $\mu = 53.7$, $\sigma = 4.7$, $\text{min} = 36.9$ \end{tabular}}}}}%
    \put(0.012,0.16){\color[rgb]{0,0,0}\rotatebox{90}{\makebox(0,0)[lt]{\lineheight{1.25}\smash{\begin{tabular}[t]{l}Capacity (bpcu)\end{tabular}}}}}%
    \put(0.367,-0.01){\color[rgb]{0,0,0}\makebox(0,0)[lt]{\lineheight{1.25}\smash{\begin{tabular}[t]{l}Transmit distance\end{tabular}}}}%
  \end{picture}%
\endgroup%

%% file: figures/jointTxRxULAs_16_4_v3.pdf_tex
%% Creator: Inkscape 1.2.1 (9c6d41e410, 2022-07-14), www.inkscape.org
%% Accompanies image file 'jointTxRxULAs_16_4_v3.pdf' (pdf, eps, ps)
\begingroup%
  \makeatletter%
  \providecommand\color[2][]{%
    \errmessage{(Inkscape) Color is used for the text in Inkscape, but the package 'color.sty' is not loaded}%
    \renewcommand\color[2][]{}%
  }%
  \providecommand\transparent[1]{%
    \errmessage{(Inkscape) Transparency is used (non-zero) for the text in Inkscape, but the package 'transparent.sty' is not loaded}%
    \renewcommand\transparent[1]{}%
  }%
  \providecommand\rotatebox[2]{#2}%
  \newcommand*\fsize{\dimexpr\f@size pt\relax}%
  \newcommand*\lineheight[1]{\fontsize{\fsize}{#1\fsize}\selectfont}%
  \ifx\svgwidth\undefined%
    \setlength{\unitlength}{2500.13470459bp}%
    \ifx\svgscale\undefined%
      \relax%
    \else%
      \setlength{\unitlength}{\unitlength * \real{\svgscale}}%
    \fi%
  \else%
    \setlength{\unitlength}{\svgwidth}%
  \fi%
  \global\let\svgwidth\undefined%
  \global\let\svgscale\undefined%
  \makeatother%
  \begin{picture}(1,0.58864343)%
    \lineheight{1}%
    \setlength\tabcolsep{0pt}%
    \put(0,0){\includegraphics[width=\unitlength,page=1]{jointTxRxULAs_16_4_v3.pdf}}%
    \put(0.012,0.18){\color[rgb]{0,0,0}\rotatebox{90}{\makebox(0,0)[lt]{\lineheight{1.25}\smash{\begin{tabular}[t]{l}Capacity (bpcu)\end{tabular}}}}}%
    \put(0.305,-0.0035){\color[rgb]{0,0,0}\makebox(0,0)[lt]{\lineheight{1.25}\smash{\begin{tabular}[t]{l}Transmit distance $\boldsymbol{d}$ (m)\end{tabular}}}}%
    \put(0,0){\includegraphics[width=\unitlength,page=2]{jointTxRxULAs_16_4_v3.pdf}}%
    \put(0.25538134,0.1912){\color[rgb]{0,0,0}\makebox(0,0)[lt]{\lineheight{1.25}\smash{\footnotesize{\begin{tabular}[t]{l}CO NULAs, $\mu=24.2, \, \sigma=2.2, \, \text{min}=16$\end{tabular}}}}}%
    \put(0.25538134,0.144){\color[rgb]{0,0,0}\makebox(0,0)[lt]{\lineheight{1.25}\smash{\footnotesize{\begin{tabular}[t]{l}CO+RR NULAs, $\mu=22.5, \, \sigma=1.1, \, \text{min}=21.1$\end{tabular}}}}}%
    \put(0,0){\includegraphics[width=\unitlength,page=3]{jointTxRxULAs_16_4_v3.pdf}}%
    \put(0.25538134,0.097){\color[rgb]{0,0,0}\makebox(0,0)[lt]{\lineheight{1.25}\smash{\footnotesize{\begin{tabular}[t]{l}ES NULAs, $\mu=23.5, \, \sigma=0.7, \, \text{min}=21.5$\end{tabular}}}}}%
    \put(0,0){\includegraphics[width=\unitlength,page=4]{jointTxRxULAs_16_4_v3.pdf}}%
    \put(0.25538134,0.238){\color[rgb]{0,0,0}\makebox(0,0)[lt]{\lineheight{1.25}\smash{\footnotesize{\begin{tabular}[t]{l}ULAs, $\mu=23.7, \, \sigma=3.4, \, \text{min}=8.8$\end{tabular}}}}}%
    \put(0,0){\includegraphics[width=\unitlength,page=5]{jointTxRxULAs_16_4_v3.pdf}}%
  \end{picture}%
\endgroup%

%% file: figures/baselineULA16.pdf_tex
%% Creator: Inkscape 1.2.1 (9c6d41e410, 2022-07-14), www.inkscape.org
%% Accompanies image file 'baselineULA16.pdf' (pdf, eps, ps)
\begingroup%
  \makeatletter%
  \providecommand\color[2][]{%
    \errmessage{(Inkscape) Color is used for the text in Inkscape, but the package 'color.sty' is not loaded}%
    \renewcommand\color[2][]{}%
  }%
  \providecommand\transparent[1]{%
    \errmessage{(Inkscape) Transparency is used (non-zero) for the text in Inkscape, but the package 'transparent.sty' is not loaded}%
    \renewcommand\transparent[1]{}%
  }%
  \providecommand\rotatebox[2]{#2}%
  \newcommand*\fsize{\dimexpr\f@size pt\relax}%
  \newcommand*\lineheight[1]{\fontsize{\fsize}{#1\fsize}\selectfont}%
  \ifx\svgwidth\undefined%
    \setlength{\unitlength}{1282.57595273bp}%
    \ifx\svgscale\undefined%
      \relax%
    \else%
      \setlength{\unitlength}{\unitlength * \real{\svgscale}}%
    \fi%
  \else%
    \setlength{\unitlength}{\svgwidth}%
  \fi%
  \global\let\svgwidth\undefined%
  \global\let\svgscale\undefined%
  \makeatother%
  \begin{picture}(1,0.02454844)%
    \lineheight{1}%
    \setlength\tabcolsep{0pt}%
    \put(0,0){\includegraphics[width=\unitlength,page=1]{baselineULA16.pdf}}%
  \end{picture}%
\endgroup%

%% file: figures/CO_NULA_Tx_16.pdf_tex
%% Creator: Inkscape 1.2.1 (9c6d41e410, 2022-07-14), www.inkscape.org
%% Accompanies image file 'CO_NULA_Tx_16.pdf' (pdf, eps, ps)
\begingroup%
  \makeatletter%
  \providecommand\color[2][]{%
    \errmessage{(Inkscape) Color is used for the text in Inkscape, but the package 'color.sty' is not loaded}%
    \renewcommand\color[2][]{}%
  }%
  \providecommand\transparent[1]{%
    \errmessage{(Inkscape) Transparency is used (non-zero) for the text in Inkscape, but the package 'transparent.sty' is not loaded}%
    \renewcommand\transparent[1]{}%
  }%
  \providecommand\rotatebox[2]{#2}%
  \newcommand*\fsize{\dimexpr\f@size pt\relax}%
  \newcommand*\lineheight[1]{\fontsize{\fsize}{#1\fsize}\selectfont}%
  \ifx\svgwidth\undefined%
    \setlength{\unitlength}{1282.57595273bp}%
    \ifx\svgscale\undefined%
      \relax%
    \else%
      \setlength{\unitlength}{\unitlength * \real{\svgscale}}%
    \fi%
  \else%
    \setlength{\unitlength}{\svgwidth}%
  \fi%
  \global\let\svgwidth\undefined%
  \global\let\svgscale\undefined%
  \makeatother%
  \begin{picture}(1,0.02454844)%
    \lineheight{1}%
    \setlength\tabcolsep{0pt}%
    \put(0,0){\includegraphics[width=\unitlength,page=1]{CO_NULA_Tx_16.pdf}}%
  \end{picture}%
\endgroup%

%% file: figures/CORR_NULA_Tx_16.pdf_tex
%% Creator: Inkscape 1.2.1 (9c6d41e410, 2022-07-14), www.inkscape.org
%% Accompanies image file 'CORR_NULA_Tx_16.pdf' (pdf, eps, ps)
\begingroup%
  \makeatletter%
  \providecommand\color[2][]{%
    \errmessage{(Inkscape) Color is used for the text in Inkscape, but the package 'color.sty' is not loaded}%
    \renewcommand\color[2][]{}%
  }%
  \providecommand\transparent[1]{%
    \errmessage{(Inkscape) Transparency is used (non-zero) for the text in Inkscape, but the package 'transparent.sty' is not loaded}%
    \renewcommand\transparent[1]{}%
  }%
  \providecommand\rotatebox[2]{#2}%
  \newcommand*\fsize{\dimexpr\f@size pt\relax}%
  \newcommand*\lineheight[1]{\fontsize{\fsize}{#1\fsize}\selectfont}%
  \ifx\svgwidth\undefined%
    \setlength{\unitlength}{1282.57595273bp}%
    \ifx\svgscale\undefined%
      \relax%
    \else%
      \setlength{\unitlength}{\unitlength * \real{\svgscale}}%
    \fi%
  \else%
    \setlength{\unitlength}{\svgwidth}%
  \fi%
  \global\let\svgwidth\undefined%
  \global\let\svgscale\undefined%
  \makeatother%
  \begin{picture}(1,0.02454844)%
    \lineheight{1}%
    \setlength\tabcolsep{0pt}%
    \put(0,0){\includegraphics[width=\unitlength,page=1]{CORR_NULA_Tx_16.pdf}}%
  \end{picture}%
\endgroup%

%% file: figures/CORR_NULA_Rx_16.pdf_tex
%% Creator: Inkscape 1.2.1 (9c6d41e410, 2022-07-14), www.inkscape.org
%% Accompanies image file 'CORR_NULA_Rx_16.pdf' (pdf, eps, ps)
\begingroup%
  \makeatletter%
  \providecommand\color[2][]{%
    \errmessage{(Inkscape) Color is used for the text in Inkscape, but the package 'color.sty' is not loaded}%
    \renewcommand\color[2][]{}%
  }%
  \providecommand\transparent[1]{%
    \errmessage{(Inkscape) Transparency is used (non-zero) for the text in Inkscape, but the package 'transparent.sty' is not loaded}%
    \renewcommand\transparent[1]{}%
  }%
  \providecommand\rotatebox[2]{#2}%
  \newcommand*\fsize{\dimexpr\f@size pt\relax}%
  \newcommand*\lineheight[1]{\fontsize{\fsize}{#1\fsize}\selectfont}%
  \ifx\svgwidth\undefined%
    \setlength{\unitlength}{1282.57595273bp}%
    \ifx\svgscale\undefined%
      \relax%
    \else%
      \setlength{\unitlength}{\unitlength * \real{\svgscale}}%
    \fi%
  \else%
    \setlength{\unitlength}{\svgwidth}%
  \fi%
  \global\let\svgwidth\undefined%
  \global\let\svgscale\undefined%
  \makeatother%
  \begin{picture}(1,0.02454844)%
    \lineheight{1}%
    \setlength\tabcolsep{0pt}%
    \put(0,0){\includegraphics[width=\unitlength,page=1]{CORR_NULA_Rx_16.pdf}}%
  \end{picture}%
\endgroup%

%% file: figures/ES_NULA_Tx_16.pdf_tex
%% Creator: Inkscape 1.2.1 (9c6d41e410, 2022-07-14), www.inkscape.org
%% Accompanies image file 'ES_NULA_Tx_16.pdf' (pdf, eps, ps)
\begingroup%
  \makeatletter%
  \providecommand\color[2][]{%
    \errmessage{(Inkscape) Color is used for the text in Inkscape, but the package 'color.sty' is not loaded}%
    \renewcommand\color[2][]{}%
  }%
  \providecommand\transparent[1]{%
    \errmessage{(Inkscape) Transparency is used (non-zero) for the text in Inkscape, but the package 'transparent.sty' is not loaded}%
    \renewcommand\transparent[1]{}%
  }%
  \providecommand\rotatebox[2]{#2}%
  \newcommand*\fsize{\dimexpr\f@size pt\relax}%
  \newcommand*\lineheight[1]{\fontsize{\fsize}{#1\fsize}\selectfont}%
  \ifx\svgwidth\undefined%
    \setlength{\unitlength}{1282.57595273bp}%
    \ifx\svgscale\undefined%
      \relax%
    \else%
      \setlength{\unitlength}{\unitlength * \real{\svgscale}}%
    \fi%
  \else%
    \setlength{\unitlength}{\svgwidth}%
  \fi%
  \global\let\svgwidth\undefined%
  \global\let\svgscale\undefined%
  \makeatother%
  \begin{picture}(1,0.02454844)%
    \lineheight{1}%
    \setlength\tabcolsep{0pt}%
    \put(0,0){\includegraphics[width=\unitlength,page=1]{ES_NULA_Tx_16.pdf}}%
  \end{picture}%
\endgroup%

%% file: figures/ES_NULA_Rx_16.pdf_tex
%% Creator: Inkscape 1.2.1 (9c6d41e410, 2022-07-14), www.inkscape.org
%% Accompanies image file 'ES_NULA_Rx_16.pdf' (pdf, eps, ps)
\begingroup%
  \makeatletter%
  \providecommand\color[2][]{%
    \errmessage{(Inkscape) Color is used for the text in Inkscape, but the package 'color.sty' is not loaded}%
    \renewcommand\color[2][]{}%
  }%
  \providecommand\transparent[1]{%
    \errmessage{(Inkscape) Transparency is used (non-zero) for the text in Inkscape, but the package 'transparent.sty' is not loaded}%
    \renewcommand\transparent[1]{}%
  }%
  \providecommand\rotatebox[2]{#2}%
  \newcommand*\fsize{\dimexpr\f@size pt\relax}%
  \newcommand*\lineheight[1]{\fontsize{\fsize}{#1\fsize}\selectfont}%
  \ifx\svgwidth\undefined%
    \setlength{\unitlength}{1282.57595273bp}%
    \ifx\svgscale\undefined%
      \relax%
    \else%
      \setlength{\unitlength}{\unitlength * \real{\svgscale}}%
    \fi%
  \else%
    \setlength{\unitlength}{\svgwidth}%
  \fi%
  \global\let\svgwidth\undefined%
  \global\let\svgscale\undefined%
  \makeatother%
  \begin{picture}(1,0.02454844)%
    \lineheight{1}%
    \setlength\tabcolsep{0pt}%
    \put(0,0){\includegraphics[width=\unitlength,page=1]{ES_NULA_Rx_16.pdf}}%
  \end{picture}%
\endgroup%

%% file: figures/jointTxRxULAs_16_4_sota_v2.pdf_tex
%% Creator: Inkscape 1.2.1 (9c6d41e410, 2022-07-14), www.inkscape.org
%% Accompanies image file 'jointTxRxULAs_16_4_sota_v2.pdf' (pdf, eps, ps)
\begingroup%
  \makeatletter%
  \providecommand\color[2][]{%
    \errmessage{(Inkscape) Color is used for the text in Inkscape, but the package 'color.sty' is not loaded}%
    \renewcommand\color[2][]{}%
  }%
  \providecommand\transparent[1]{%
    \errmessage{(Inkscape) Transparency is used (non-zero) for the text in Inkscape, but the package 'transparent.sty' is not loaded}%
    \renewcommand\transparent[1]{}%
  }%
  \providecommand\rotatebox[2]{#2}%
  \newcommand*\fsize{\dimexpr\f@size pt\relax}%
  \newcommand*\lineheight[1]{\fontsize{\fsize}{#1\fsize}\selectfont}%
  \ifx\svgwidth\undefined%
    \setlength{\unitlength}{2496.88531494bp}%
    \ifx\svgscale\undefined%
      \relax%
    \else%
      \setlength{\unitlength}{\unitlength * \real{\svgscale}}%
    \fi%
  \else%
    \setlength{\unitlength}{\svgwidth}%
  \fi%
  \global\let\svgwidth\undefined%
  \global\let\svgscale\undefined%
  \makeatother%
  \begin{picture}(1,0.59957351)%
    \lineheight{1}%
    \setlength\tabcolsep{0pt}%
    \put(0,0){\includegraphics[width=\unitlength,page=1]{jointTxRxULAs_16_4_sota_v2.pdf}}%
    \put(0.24077612,0.21){\color[rgb]{0,0,0}\makebox(0,0)[lt]{\lineheight{1.25}\smash{\footnotesize{\begin{tabular}[t]{l}CO+RR NULAs, $\mu=22.5, \, \sigma=1.1, \, \text{min}=21.1$\end{tabular}}}}}%
    \put(0.24077612,0.165){\color[rgb]{0,0,0}\makebox(0,0)[lt]{\lineheight{1.25}\smash{\footnotesize{\begin{tabular}[t]{l}[10] NULAs, $\mu=24, \, \sigma=2.3, \, \text{min}=13.8$\end{tabular}}}}}%
    \put(0.24077612,0.11932642){\color[rgb]{0,0,0}\makebox(0,0)[lt]{\lineheight{1.25}\smash{\footnotesize{\begin{tabular}[t]{l}[11] NULAs, $\mu=22.2, \, \sigma=2.3, \, \text{min}=17.9$\end{tabular}}}}}%
    \put(0.008,0.184){\color[rgb]{0,0,0}\rotatebox{90}{\makebox(0,0)[lt]{\lineheight{1.25}\smash{\begin{tabular}[t]{l}Capacity (bpcu)\end{tabular}}}}}%
    \put(0.3,0.000){\color[rgb]{0,0,0}\makebox(0,0)[lt]{\lineheight{1.25}\smash{\begin{tabular}[t]{l}Transmit distance $\boldsymbol{d}$ (m)\end{tabular}}}}%
  \end{picture}%
\endgroup%

%% file: figures/jointTxRxULAs_16_4_multiNf_v5.pdf_tex
%% Creator: Inkscape 1.2.1 (9c6d41e410, 2022-07-14), www.inkscape.org
%% Accompanies image file 'jointTxRxULAs_16_4_multiNf_v5.pdf' (pdf, eps, ps)
\begingroup%
  \makeatletter%
  \providecommand\color[2][]{%
    \errmessage{(Inkscape) Color is used for the text in Inkscape, but the package 'color.sty' is not loaded}%
    \renewcommand\color[2][]{}%
  }%
  \providecommand\transparent[1]{%
    \errmessage{(Inkscape) Transparency is used (non-zero) for the text in Inkscape, but the package 'transparent.sty' is not loaded}%
    \renewcommand\transparent[1]{}%
  }%
  \providecommand\rotatebox[2]{#2}%
  \newcommand*\fsize{\dimexpr\f@size pt\relax}%
  \newcommand*\lineheight[1]{\fontsize{\fsize}{#1\fsize}\selectfont}%
  \ifx\svgwidth\undefined%
    \setlength{\unitlength}{2442.23309326bp}%
    \ifx\svgscale\undefined%
      \relax%
    \else%
      \setlength{\unitlength}{\unitlength * \real{\svgscale}}%
    \fi%
  \else%
    \setlength{\unitlength}{\svgwidth}%
  \fi%
  \global\let\svgwidth\undefined%
  \global\let\svgscale\undefined%
  \makeatother%
  \begin{picture}(1,0.60131131)%
    \lineheight{1}%
    \setlength\tabcolsep{0pt}%
    \put(0,0){\includegraphics[width=\unitlength,page=1]{jointTxRxULAs_16_4_multiNf_v5.pdf}}%
    \put(0.3,-0.02){\color[rgb]{0,0,0}\makebox(0,0)[lt]{\lineheight{1.25}\smash{\begin{tabular}[t]{l}Uniform grid spacing $d_{\text{t,f}}$\end{tabular}}}}%
    \put(0.065,0.033){\color[rgb]{0,0,0}\makebox(0,0)[lt]{\lineheight{1.25}\smash{\begin{tabular}[t]{l}$d_{\text{t}}/2$\end{tabular}}}}%
    \put(0.186,0.033){\color[rgb]{0,0,0}\makebox(0,0)[lt]{\lineheight{1.25}\smash{\begin{tabular}[t]{l}$d_{\text{t}}/4$\end{tabular}}}}%
    \put(0.43,0.033){\color[rgb]{0,0,0}\makebox(0,0)[lt]{\lineheight{1.25}\smash{\begin{tabular}[t]{l}$d_{\text{t}}/8$\end{tabular}}}}%
    \put(0.915,0.033){\color[rgb]{0,0,0}\makebox(0,0)[lt]{\lineheight{1.25}\smash{\begin{tabular}[t]{l}$d_{\text{t}}/16$\end{tabular}}}}%
    \put(0.333,0.033){\color[rgb]{0,0,0}\makebox(0,0)[lt]{\lineheight{1.25}\smash{\begin{tabular}[t]{l}$d_{\text{t}}/5$\end{tabular}}}}%
    \put(0.014,0.14){\color[rgb]{0,0,0}\rotatebox{90}{\makebox(0,0)[lt]{\lineheight{1.25}\smash{\begin{tabular}[t]{l}Min. capacity (bpcu)\end{tabular}}}}}%
  \end{picture}%
\endgroup%